\documentclass[journal]{IEEEtran}
\usepackage{graphicx}
\usepackage[dvipsnames]{xcolor}
\usepackage{algpseudocode}
\usepackage{amssymb}
\usepackage{algorithm}

\usepackage{xcolor}
\usepackage{enumerate}% http://ctan.org/pkg/enumerate

\usepackage{cite}
\usepackage{caption}
\ifCLASSINFOpdf
\else
\fi
\begin{document}
\title{A Lightweight Security Solution for Mitigation of Hatchetman Attack in RPL-based 6LoWPAN}
\author{Girish Sharma$^{*,a,b}$, Jyoti Grover{$^a$},~\IEEEmembership{Senior Member,~IEEE}, Abhishek Verma{$^c$},~\IEEEmembership{Member,~IEEE,}  % <-this % stops a space
\newline $^{a}$ Malaviya National Institute of Technology Jaipur, JLN Marg, Jaipur, Rajasthan, India 302017
\newline $^{b}$ Manipal University Jaipur, Dehmi Kalan, Jaipur, Rajasthan, India 303007
\newline $^{c}$ Department of Information Technology, Babasaheb Bhimrao Ambedkar University, Uttar Pradesh, India
\thanks{G. Sharma, J. Grover are with the Department of Computer Science \& Engineering, Malaviya National Insitiute of Technology Jaipur, Rajasthan, India, 302017. G. Sharma is also with Manipal University, Dehmi Kalan, Jaipur, Rajasthan 303007. A. Verma is with the Department of Information Technology, Babasaheb Bhimrao Ambedkar University, Lucknow, Uttar Pradesh, India,  e-mails: jgrover.cse@mnit.ac.in, 2020rcp9012@mnit.ac.in, abhiverma@iiitdmj.ac.in, }% <-this % stops a space
% \thanks{J. Doe and J. Doe are with Anonymous University.}% <-this % stops a space
% \thanks{Manuscript received April 19, 2005; revised August 26, 2015.}
\thanks{*Corresponding Author}
}

%\markboth{Journal of \LaTeX\ Class Files,~Vol.~, No.~, December~2023}%

\maketitle

\begin{abstract}
In recent times, the Internet of Things~(IoT) has a significant rise in industries, and we live in the era of Industry 4.0, where each device is connected to the Internet from small to big.  These devices are Artificial Intelligence~(AI) enabled and are capable of perspective analytics. By 2023, it's anticipated that over 14 billion smart devices will be available on the Internet. These applications operate in a wireless environment where memory, power, and other resource limitations apply to the nodes. In addition, the conventional routing method is ineffective in networks with limited resource devices, lossy links, and slow data rates. Routing Protocol for Low Power and Lossy Networks~(RPL), a new routing protocol for such networks, was proposed by the IETF's ROLL group. RPL operates in two modes: Storing and Non-Storing. In Storing mode, each node have the information to reach to other node. In Non-Storing mode, the routing information lies with the root node only. The attacker may exploit the Non-Storing feature of the RPL. When the root node transmits User Datagram Protocol~(UDP) or control message packet to the child nodes, the routing information is stored in the extended header of the IPv6 packet. The attacker may modify the address from the source routing header which leads to Denial of Service~(DoS) attack. This attack is RPL specific which is known as Hatchetman attack. This paper shows significant degradation in terms of network performance when an attacker exploits this feature. We also propose a lightweight mitigation of Hatchetman attack using game theoretic approach to detect the Hatchetman attack in IoT.

\footnote{https://ieeexplore.ieee.org/abstract/document/10469481/}

\end{abstract}


\IEEEpeerreviewmaketitle

\section{Introduction}\label{Introduction}

\IEEEPARstart{I}{}nternet of Things is one of the key technology for Industry 4.0. IoT and Industrial IoT~(IIoT) enable many applications in the consumer world~\cite{khan2021role}. It is expected that around 27 billion IoT devices will be connected to the Internet~\cite{forecast} by the year 2025. IoT is connecting different devices ranging from Radio Frequency Identification~(RFID), smart grids, Wireless Sensor Network~(WSN) to the Internet. The different devices connected to the Internet makes IoT networks heterogeneous~\cite{sain2017survey} and it uses different protocol standards as compared to the traditional protocol stack. One of the protocols that an IoT protocol stack uses is 6LoWPAN~\cite{wallgren2013routing} which works as an adaptation layer. This layer supports the constrained nodes and the networks with low data rates. 6LoWPAN was designed to optimize the IPv6 packets over the lossy and low data rate constrained networks such as IEEE 802.15.4. The adaptation layer handles header compression, fragmentation and mesh addressing i.e. mutiple hops forwarding of IPv6 packets.The IETF Routing Over Low-Power and Lossy Networks~(ROLL) working group  proposed a new protocol commonly known as Routing Protocol for Low Power and Lossy Networks~(RPL) for the constrained networks. RPL, which is based on distance vector routing protocol optimizes the resource requirements along the routing path.  \par Applications for the Internet of Things are growing every day, and RPL was created specifically for them. So there is a need to address different security aspects of RPL. Routing attacks on RPL have gained lot of attention by researchers since some are inherited from WSN and some are RPL specific. One such attack is Hatchetman attack which has drastic impact in terms of throughput and Packet Delivery Ratio~(PDR). This attack was first addressed by Cong Pu \textit{et al.} in their article~\cite{pu2018hatchetman}. In their paper, the performance evaluation is carried out in terms of PDR, Energy consumption by changing the source routing header. Latency in packet delivery, energy usage, and throughput are all impacted by a hatchetman attack. During packet forwarding, this attack dynamically changes the source routing header, leading to a situation where legitimate nodes are unable to forward the packet. Consequently, this gives rise to a denial of service attack. RPL supports non-storing mode for downward packet delivery which is useful for LLNs because of memory constraint~\cite{oh2018hybrid}. The root node stores the IPv6 address of each hop between the source and destination. The attacker may take advantage of this information and fills the Source Route Header~(SRH) with the fake IPv6 address. This results in the propagation of ICMPv6 error messages to the SRH generator. In this paper we analyze the impact of hatchetman attack on the network and also proposes the mitigation of the attack for RPL based LLNs.

 \subsection{\textbf{Contributions}}
 \par The contributions of our paper are listed below:
 \begin{enumerate}
     \item Implementation and analysis of the hatchetman attack in RPL-based static and mobile IoT networks. 
     \item Lightweight mitigation of hatchetman attack using game theoretic approach.
   
 \end{enumerate}

\subsection{\textbf{Organization}}
\par The remaining paper is organized as follows: Section~\ref{Background} overview of RPL protocol. This section also outlines the functioning of the hatchetman attack. The implementation of hatchetman attack is discussed in section~\ref{attackImplementation}. In the section~\ref{systemmodel}, we discuss the detection approach and mathematical formulation of the performance parameters. We also propose an algorithm for the detection of the attack in the section~\ref{solutionapproach}. At last, Section~\ref{CFW} concludes some useful insights and presents future study directions. 

\section{Background}\label{Background}
% This section discusses the RPL protocol and Hatchetman Attack
\subsection{Overview of RPL Protocol} \label{RPL-Overview}
Using the distance vector routing concept, RPL sets up networks as Directed Acyclic Graphs~(DAGs) and serves Low Power Lossy Networks~(LLN). The DAG topology has one or more root nodes, which is generally a border router that connects the outside world to the other nodes in the DAG. RPL works on two modes to implement the resource-constrained nature of the nodes: a) Storing  b) Non-Storing. a) In storing mode, each node has a complete path to route a packet to the sink node or the other nodes. b) In non-storing mode, only the border router has complete information for the nodes, i.e., the downward path. In this mode, the nodes only have a parents' list so they can forward the packets to the border router.  
\par RPL, based on the destination-oriented source routing protocol, implements five control messages: a) \textit{DODAG Information Object~(DIO)}: This is used to maintain and configure the upward routes. The nodes listen to the DIO messages to configure as per the changes in the topology and determines the parent and the route to the sink using the information provided in DIO message~~\cite{kumar2022addressing}.  
b) \textit{DODAG Information Solicitation~(DIS)}: DIS is used to probe the network when the node wants to join the DODAG. If the DIS was sent as a unicast, the receiving node would transmit the DIO message containing the DODAG configuration. In the case of multicast, the receiving nodes would reset the trickle timer and broadcast the DIO messages.
c,d) \textit{Destination Advertisement Object~(DAO) and Acknowledgment~(DAO-
ACK) Messages}:
The nodes maintain the downward routes by sending the parent information toward the border router using the DAO message. The root maintains this information to send the datagram to the nodes by determining the source route. The receiver responds to the sender by sending DAO-ACK depending on the flag field of the DAO message~\cite{RPLNutshell}.
e) Consistency Check~(CC) messages: These are used for coordinating time and security measurements between any two nodes. These messages are only available in RPL' secure mode.
\par RPL's objective function~(OF) is used to determine the node's rank.  Various OFs in RPL include ETX Objective function~(ETXOF), Minimum Rank with Hysteresis Objective Function~(MRHOF), and Objective Function Zero~(OF0). RFC 6551 defines some of the routing metrics such as Expected Transmission Count~(ETX), Hop Count, Latency, Link Quality Level, Node Energy and Throughput. The rank of a node determines how close it is to the root node. Rank helps the DODAG to remove count-to-infinity and loops in the network. The node can decrease its rank if it finds a lower-cost route to the root~\cite{thulasiraman2019lightweight}.
\par In summary, RPL is a new proposed standard that facilitates applications based on the IPv6 protocol. With the advent of IoT as a new industry standard, RPL has become one of the most important network layer protocols in sensor networks due to the resource-constrained nature of nodes.

\subsection{Hatchetman Attack in RPL} 

\label{Hatchetman Understanding}

This section discusses implementation of Hatchetman attack using the \textit{Contiki} Operating System. RPL protocol is based on distance vector source routing mechanism. The attacker can exploit the extended header of IPv6. IPv6 packet has fixed base header of size 40 bytes. The extension headers contains different options depending on the requirement. The source fills all these extension header information~\cite{sharma2022analysis}. One such option used in case of source routing is \textbf{Routing Extension Header~(REH)}. The source node stores the complete routing path. The intermediate nodes transmit the packet to the subsequent destination, and as the packet follows its path, it ultimately reaches the intended destination. 

\begin{figure}[h]
    \centering
    \includegraphics[width=3.6in,height=4.1in,keepaspectratio]{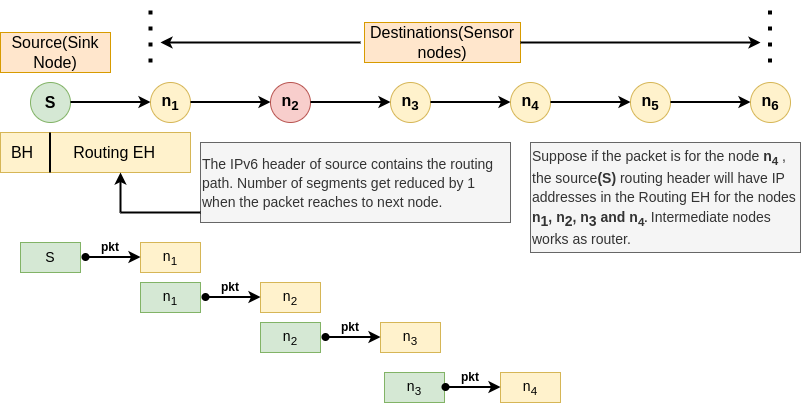}
    \caption{Normal Scenario: Packet Delivers Successfully~(Downward PDR) }
    \label{Normal-Scenario}
\end{figure}

\begin{figure}[h]
    \centering
    \includegraphics[width=3.1in,height=4.0in,keepaspectratio]{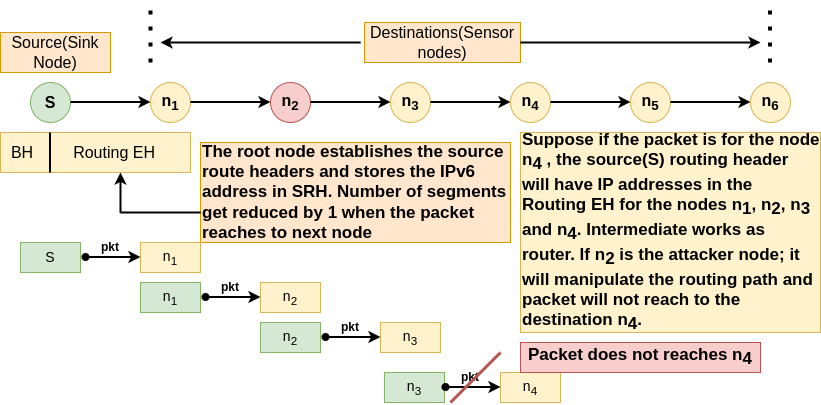}
    \caption{\textbf{Attack Scenario: Packets do not reach node $n_4$ onwards}}
    \label{Attack-Scenario}
\end{figure}

Figure~\ref{Normal-Scenario} shows a simple line topology. The source is connected to all the nodes through the intermediate nodes. The \textbf{strict source routing } mechanism sends the packet from the source to destination when there is no attack. The \textbf{Segments Left} field of the SRH shows that number of intermediate nodes to be visited to reach to the destination. Hatchetman attack scenario is shown in the Fig.~\ref{Attack-Scenario} But if an attacker changes the IPv6 address in the source route path with an unreachable IPv6 address, the packet will not reach the destination, and this will lead to the Denial of Service Attack~(DoS).

This attack reduces the downward PDR and also increases the ICMPv6 error messages as shown in section~\ref{attackImplementation}. The node's energy consumption rises as it repeatedly tries to send the packet on to the next hop.  

\section{\textbf{Hatchetman Attack Implementation}}\label{attackImplementation}
This section shows the impact of the hatchetman attack on static and mobile RPL based IoT. In contrast to the vast majority of research, which focus on static networks, this paper evaluates performance metrics for mobile IoT.
\subsection{Experimental Setup}
We use cross level COOJA simulator for the performance evaluation. COOJA simulator which was developed specifically for IoT supports RFC 6550, uses Contiki as underlying operating system for sensor nodes. Contiki 3.0 optimizes the IoT standards proposed in RFC 6550. Wireless technology for the media access control and physical layers in constrained networks is also supported by Contiki as it implements IEEE 802.15.4. At network layer Contiki implements RPL as the routing protocol which is based on distance vector routing protocol.
\par The Z1 platform, a low-power microcontroller, is used in the experiment since its transceiver can communicate at 2.4GHz. Z1 is compliant with IEEE 802.15.4 and can handle low data rates. 
\par We use COOJA simulator for the normal and attack scenarios simulation. It has a built-in hardware simulator called MSPsim that replicates the same binary code as sensor devices. This makes the models more realistic. We use the Z1 platform, which is a node for 6LoWPAN. In this simulation, the Unit Disc Graph Medium (UGDM) radio model is used. The test is run on a grid that was 200m x 200m and had between 10 and 50 nodes. We also use Random Waypoint Mobility Model to simulate the mobile networks where speeds of nodes varies from $1-2\;m/sec$.

\begin{table}[!hbt]
\caption{\label{setup}\textbf{Experimental simulation parameters}}
\begin{center}
\begin{tabular}{ |p{4cm}|p{3.5cm}| }
\hline
\textbf{Parameters} & \textbf{Value}\\
\hline

Grid Size 		& $200m\; X \;200 m$ \\
\hline 
Sensor Nodes 	& $10\;to\;30$\\
\hline
Gateway Nodes 	&	$1$\\
\hline
Radio Medium		& Unit Disk Graph Medium \\
\hline
Physical and Medium Access Control Layer		& IEEE 802.15.4 \\
\hline
Transmission Range		& $50\;m$ \\
\hline
Interference Range		& $100\; m$ \\
\hline
Number of Attacker nodes &	$1$ \\
\hline
Speed of Node		&   $1-2\;m/sec$ \\
\hline
Data Packet Size		& $30\;bytes$ \\
\hline
Data packet sending interval & $60\;secs$ \\
\hline
\end{tabular}
\end{center}
\end{table}

\subsection{Hatchetman Attack Algorithm}
\par To implement the attack we make changes in \textit{rpl-ext-header.c} of Contiki operating system. The attacker node modifies the IPv6 address in the Source Routing Header~(SRH), thereby preventing the next hop from successfully forwarding the packet to its intended destination.
The Next Header bits of SRH provides the type of the next header as in IPv6 Next Header field. The Routing Type value is 3 in SRH. The Segments Left field is decremented when the packet moves from one hop to next hop. The 4 bits CmprI fields stores number of prefix octets for the segments except the last one and 4 bits CmprE bits specifies the prefix octets for the last segment. The address field stores the IPv6 address of all the hops through which the packet will move towards the destination. 
\par Algorithm~\ref{attack-implementation} shows the implementation of hatchetman attack by changing \textit{rpl-ext-header.c} file of Contiki. This algorithm processes SRH as per the rfc 6554. The algorithm works as follows:
\begin{enumerate}[i]
\item The attacker node fetches the SRH and finds the index of next to next destination by checking the number of Segments Left.
     \item If the IPv6 address is available, the attacker modifies using the random() fucntion.
     \item When the packet reaches to the next hop, it can not forward the packet because of random address.
     \item The node tries to send the packet multiple times but fails and this generate ICMP error message to be propagated to SRH generator.
     \item The attack causes a denial of service attack. The critical aspect of this attack is that the attacker forwards the packet to the next hop, and it is difficult to find the attacker.
     \item This algorithm takes constant amount of time~($O(1)$) as there is no loop involved.
 \end{enumerate}

\begin{algorithm}
\flushleft
  \caption{Hatchetman Attack}
  \label{attack-implementation}
  \begin{algorithmic}[1] % The number tells where the line numbering should start
  \Comment{Processing of source routing Header by the attacker node}
  \If{$Segments\;Left==0$}
     \State Read Routing header's Next Header field and process the next header
  
    \Else {
    \State Find out $n$ how many addresses are in the Routing header\newline
    \Comment{ n = (((Hdr Ext Len * 8) - Pad - (16 - CmprE)) / (16 - CmprI)) + 1 The values are taken from source routing header}
    
    \If{$Segments\;Left > n $}
     \State Discard the packet. ICMP Error
    \Else{\State $Segments\;Left = Segments Left  - 1 $ 
    \State Compute $index$ of the next address to be visited in the Address[1..n] of SRH}

    \State Compute the $index\_Next$ of the next to next address to be visited in the Address[1..n] of SRH  
    \State Store a random address at the next to next in the vector Address[1..n]    \Comment{This causes that the attacker forwards the packet to next hop. But the next hop can not forward the packet because SRH has been illegitimately modified.}

    \State Swap the IPv6 Current Destination Address and next address computed in previous step
    \State Decrement hop limit
     \EndIf
     }
    \EndIf
    \State \textbf{end procedure} 
  \end{algorithmic}
\end{algorithm}

\begin{figure}[h]
     \centering
    \includegraphics[width=3.1in,height=4.8in,keepaspectratio]{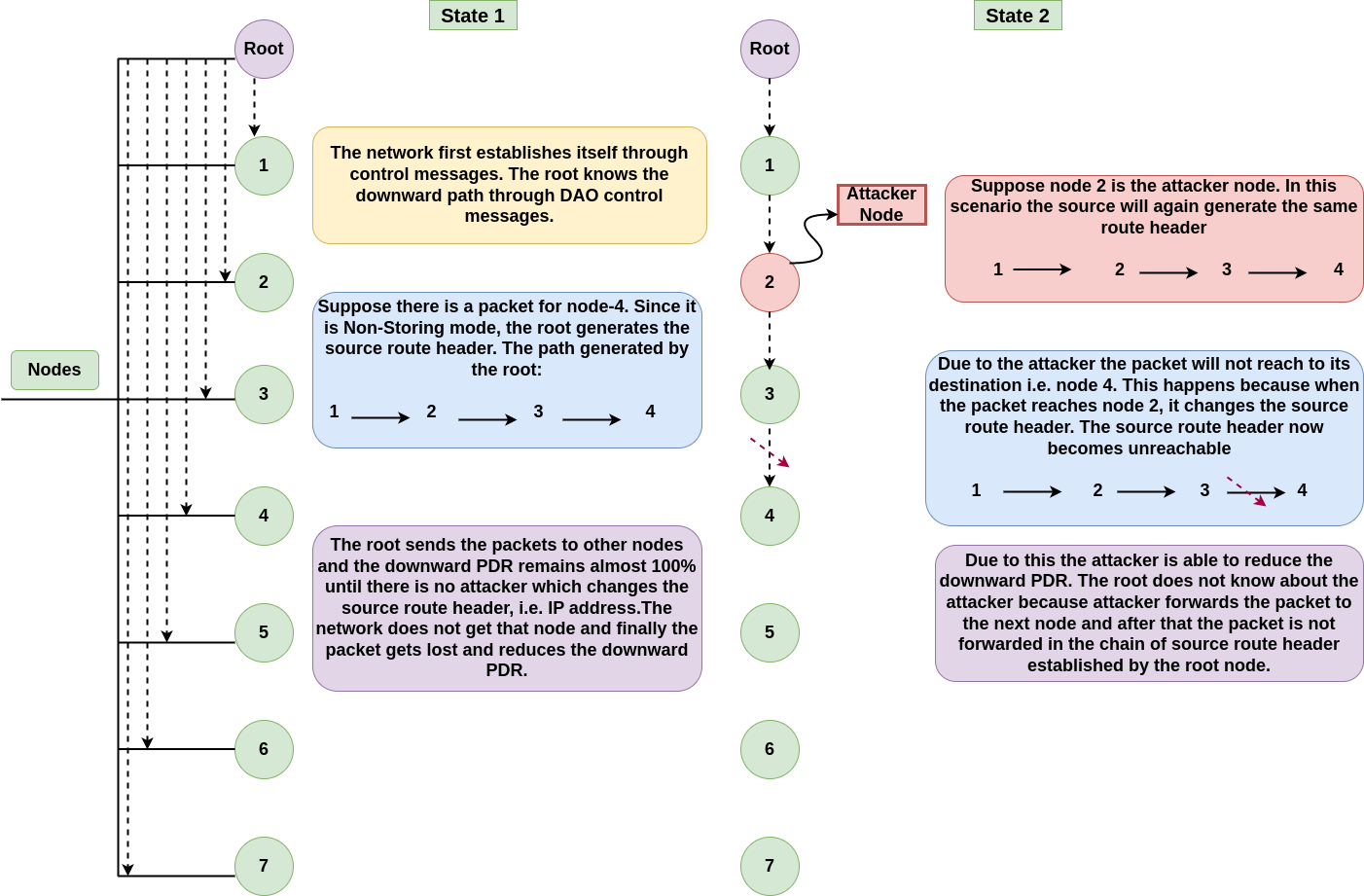}
\centering    \caption{\textbf{Hatchetman Attack Process}}
    \label{HAP}
\end{figure}

This attack process is shown in the Fig.~\ref{HAP}. If the attacker changes the SRH, the node next to it will not be able to forward the packet to the next node reducing the downward PDR. Figure~\ref{HAP} also shows that the attacker's position is vital. 

\section{\textbf{System Model}}\label{systemmodel}
\par This section shows our lightweight Non-cooperative Game Theory based approach for detecting the Hatchetman attack~\cite{han2012game}. Mathematical models of cooperation and conflict between rational, intelligent agents are the primary objective of game theory. It has played a significant role in Wireless Sensor Networks for detecting the attacker nodes~\cite{rehman2019sinkhole,saad2009coalitional}. We propose the following matix game for detection of Hatchetman attack. The game consists of strategies, players and their actions. 

\begin{table}[!hbt]
\caption{\label{GameTheory}\textbf{Notations for Hatchetman Attack Detection Game}}
\begin{center}
\begin{tabular}{ |p{5.3cm}|p{2.6cm}| }
\hline
\textbf{Description} & \textbf{Notation}\\
\hline

Number of finite nodes~(Agents)		& $n$ \\
\hline 
Strategy-1 Node forwards packet 	& $Fp$\\
\hline
Strategy-2 Node does not forward packets 	& $Dfp$\\
\hline
Payoff function of the player~(node) $i$	& $u_{i}(i,j); u_{i}: S\rightarrow \mathbb{R} $\\
\hline
Average Power Consumption at node $i$ 	& $\epsilon_i$\\
\hline
End to End Delay at node $i$ 	& $\delta_i$\\
\hline
Downward Packet Delivery Ratio node $i$ 	& $\mu_i$\\
\hline
Number of Packets	& $\mathcal{N}$\\
\hline
clock\_ticks: number of ticks	& $c_t$\\
\hline
Duration of ticks	& $c_d$\\
\hline

Current consumption	& $c_c$\\
\hline
Number of packets sent by sink to sensor node & $Si_n$\\
\hline
Number of packets received by sensor nodes	& $Sn_n$\\

\hline
\end{tabular}
\end{center}
\end{table}

\begin{figure}[h]
    \centering
    \includegraphics[width=2.9in,height=3.2in,keepaspectratio]{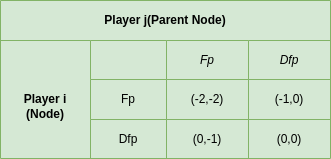}
   
    \caption{\textbf{Two Players Game for Hatchetman Attack Detection}}
 
    \label{matrix}
\end{figure}

The game can be represented by the matrix displayed in Fig.~\ref{matrix}. Rows and columns make up a game's matrix. Player's tactics is depicted by the rows and columns. Player's expected returns on various strategies are represented by rows and columns in the matrix. To understand the concept of matrix games let us assume that there are $2$ players. In non-storing mode of RPL, each node is either forwarder or the recipient of the packet. So we can model the two player matrix game as in Fig.~\ref{matrix}.
\begin{enumerate}[i]
    \item The Fig.~\ref{matrix} shows the payoff of each player~(node) depending on whether it forwards the packet or not.
    \item This matrix is maintained for each player where player $i$ represents the node itself and player $j$ is the parent of node $i$.
    \item If the node $i$ is not able to forward the packet due to modifications by the parent $j$; this signifies the payoff $(0,-1)$. 
    \item If the parent is not able to forward the packet; this situation already leads to attack for the node previous to the parent.
\end{enumerate}

Fig.~\ref{matrix} illustrates the pure strategy Nash Equilibrium (PSNE) in which both players choose strategy (Dfp, Dfp), resulting in player i being the attacker node. 

\subsection{\textbf{Mathematical Formulation}}
To solve the non-cooperating games as depicted in Fig.~\ref{matrix}, we use the dominating strategies to eliminate the rows and columns to get the solution. In our proposed approach the attacker node payoff will be the dominant player.\\
\textbf{Attacker Prediction:} Here we find the attacker mathematically by using dominant strategy. 
A strategy $s'_i \in S_i$ of a node is dominated by a strategy $s_i \in S_i$  if the payoff function of the strategy exceeds that of the node~\ref{eqn1}. 

 \begin{equation}\label{eqn1}
    u_i(s_i,s_{-i}) 	\geq  u_i(s'_i,s_{-i}),\forall s_{-i}\in S_{-i}   
\end{equation}
In the Fig.~\ref{matrix} we can see that $Fp$ is dominated by $Dfp$ for both the nodes which reduces the number of possibilities and lead to outcome of the game. In our proposed approach, we can find the attacker node by eliminating the rows non-dominating rows and columns.\\
\textbf{Delay: } Average time required for all packets to reach the destination application layer from the moment the packet is sent from the source's application. This is calculated as per equation~\ref{eqn2}
 \begin{equation}\label{eqn2}
Delay = \frac{\sum \delta_i}{\mathcal{N}}
\end{equation}
\textbf{Packet Delivery Ratio:}  This is the average number  
 \begin{equation}\label{eqn3}
\mu_i= \frac{\sum S_i}{Sn_n}
\end{equation}

 \begin{figure*}[h]
    \centering
   \includegraphics[width=5.4in,height=3.2in,keepaspectratio]{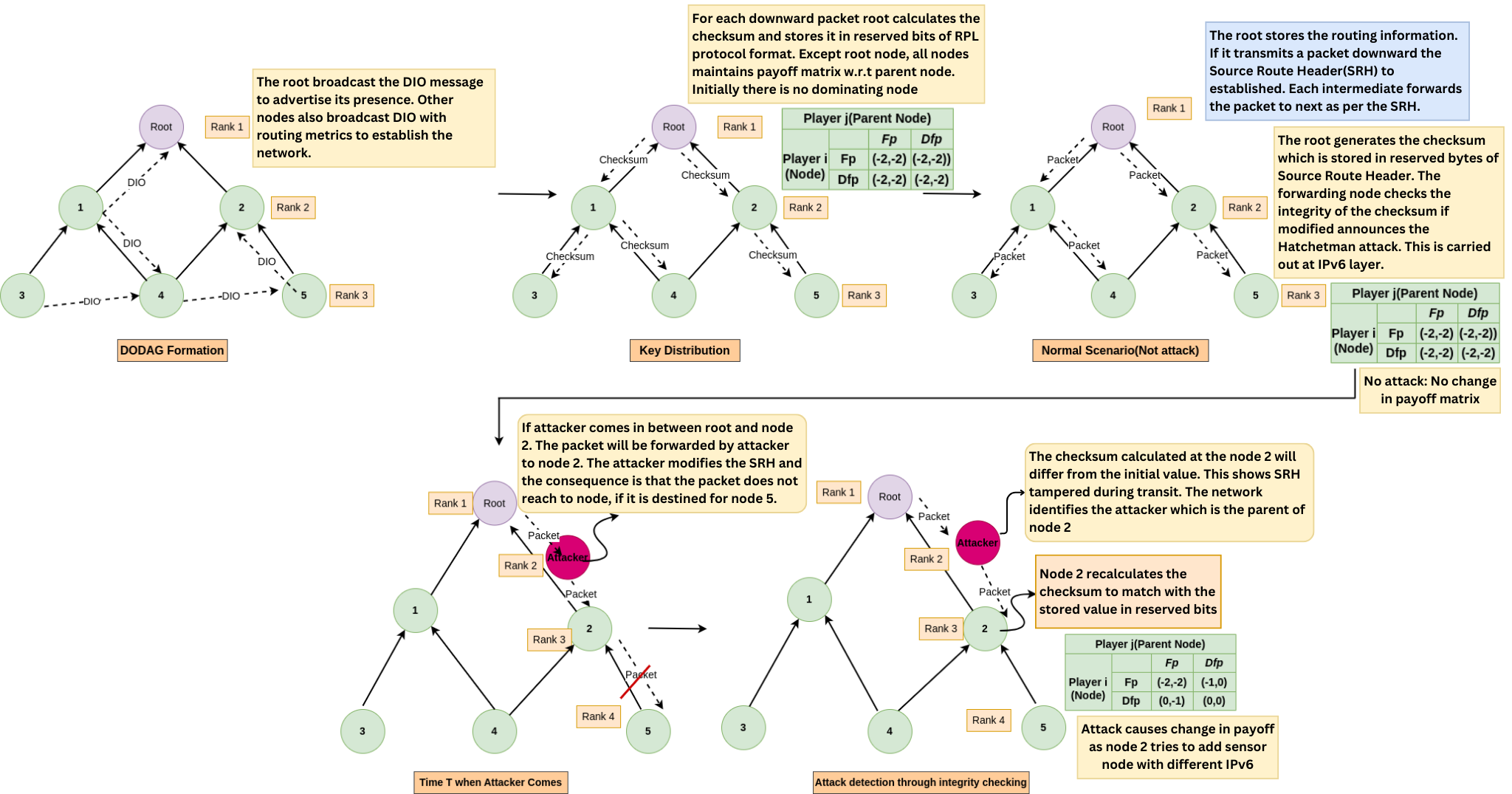}
    \caption{\textbf{Hatchetman attack detection using matrix game approach }}
 
    \label{AttackDetectionFlow}
\end{figure*}

\textbf{Average Power Consumption: }   Average of sum of power that the sensor node used. 
 \begin{equation}\label{eqn4}
 \epsilon_i= avg\_current * voltage\; mW
\end{equation}
where $voltage$ is which the system provides to the components and $avg\_current$ is $(c_t*c_c)/c_d\;mA$

The performance parameters resuts are shown in the section~\ref{solutionapproach}. 
 
\subsection{\textbf{Model Explanation} }

   \begin{figure}[h]
    \centering
    \includegraphics[width=3.2in,height=3.8in,keepaspectratio]{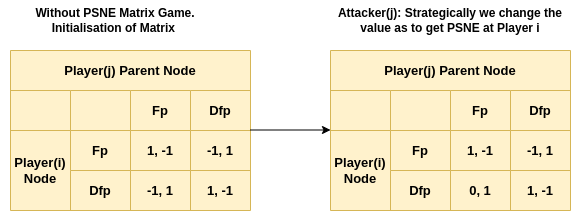}
    \caption{\textbf{How the Model Works }}
    \label{Model-Proof}
\end{figure}

The matrix game between node $i$ and its parent $j$ is depicted in Fig.~\ref{Model-Proof}. The table is initialised without any Pure Strategy Nash Equilibria (PSNE) for the matrix game. In the event of an attack, node $i$ will be capable of detecting the intrusion due to its inability to send the packet. The proposed solution for the PSNE is represented by the tuple (Dfp, Fp).  

\section{\textbf{Solution Approach}}\label{solutionapproach}
The solution to the hatchetman attack is based on the game theory. Game theoretic approach provides a good formulation for network attacks mitigation. The Algo.~\ref{alg2} shows how we can detect and mitigate the hatchetman attack in IoT. The notation $u(i,j)$ represents the payoff matrix, $B_{l}(k)$ blacklist nodes, $ch_{i}, ch_{n}$ represents the initially calculated checksum and current checksum respectively. $SN_{ip}$ is IPv6 control message to add the sensor node.

\begin{algorithm}
\caption{Hatchetman Attack Detection Process}\label{alg2}
\begin{algorithmic}[1]
\Require $u(i,j)$, SRH \Comment{Payoff Matrix}
\Ensure $B_l[k]$ \Comment{Black List Nodes}
\State \texttt{Initialize the payoff matrix at each sensor node}
\For{\texttt{each downward packet at sensor node}}
  \State{\texttt{Read SRH to get the next address}}
   \If{Address is available}
        \State \texttt{Forward the packet to next node}
    \ElsIf{$Node_i$ generates $SN_{ip}$ \& $ch_i!=ch_n$}
    \State $u(i,j) = (0,-1)$ \Comment{Modify the payoff matrix to get the dominating node}
    \EndIf    
 \EndFor
 \For{$i=0;i<2;i++$}
\For{$j=0;j<2;j++$} 
 \If{$u(i,j) == (0,1)$} 
\State \texttt{Add  $Node_j$ to $B_l[k]$ } \Comment{Attacker node}
\EndIf
\EndFor
\EndFor
\end{algorithmic}
\end{algorithm}
The Algo.~\ref{alg2} detects the attacker node by using the elegant game theory approach. The root node calculates the checksum based on the SRH and stores it in the reserved bits of RPL protocol. Each sensor node also maintains payoff matrix. When the attacker modifies the SRH as discussed in section~\ref{Background}; the sensor node generates IPv6 packet to add new neighbour in DODAG. This fake node is actually not available in the network. Apart from this, the checksum calculated will be different from the original one. Figure~\ref{AttackDetectionFlow} depicts our implementation approach for detecting attacker nodes. Our lightweight method does not rely on cryptographic techniques or MAC-based solutions~\cite{sharma2023qsec}.

\subsection{\textbf{Simulation Result}}\label{result-analysis}
This section~\ref{result-analysis} outlines the outcomes of our implementation of the Hatchetman attack on various performance indicators.
\subsubsection{\textbf{Impact on Downward PDR}}
    \begin{figure}[h]
    \centering
    \includegraphics[width=3.0in,height=4.0in,keepaspectratio]{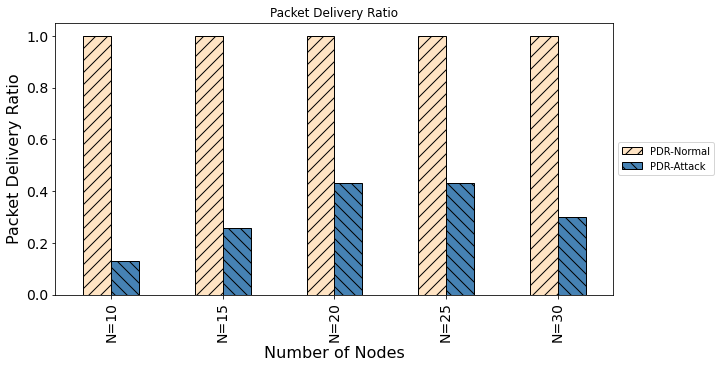}
    \caption{\textbf{Impact on Downward PDR }}
    \label{DownPDR}
\end{figure}
Fig.~\ref{DownPDR} shows impact on PDR as we increases number of nodes. For a normal scenario, PDR is $1$, but when the attacker node is placed at 1-hop distance, it reduces the PDR. 

\subsubsection{\textbf{Impact on AE2ED}} 
\begin{figure}[h]
    \centering
    \includegraphics[width=3.0in,height=4.0in,keepaspectratio]{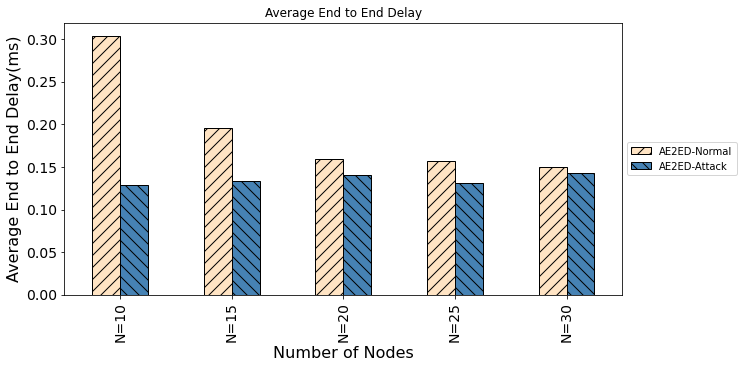}
     \caption{\textbf{Average End to End Delay with and without attack. }}
    \label{AE2ED}
\end{figure}

Figure~\ref{AE2ED} demonstrates a comparable impact on the average end-to-end delay when the number of nodes is increased. As the quantity of generated and received packets decreases, the AE2ED also decreases.

\subsubsection{\textbf{Impact on Overhead Packets}} 
\begin{figure}[h]
    \centering
    \includegraphics[width=3.0in,height=4.0in,keepaspectratio]{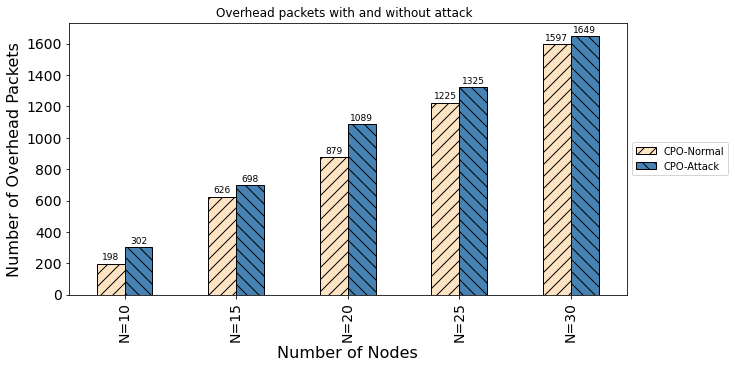}
     \caption{\textbf{Increase in Overhead Packets }}
    \label{CPO-Overhead}
\end{figure}

The attacker node also increases the overhead packets as shown in Fig.~\ref{CPO-Overhead}. These packets are IPv6 control messages and some control messages are due to the hatchetman attack when the node cannot forward the packet. 

\section{\textbf{Conclusion and Future Scope}}\label{CFW}
IoT has powered Industry 4.0, and it has significantly impacted our life. Each small to big device will be connected to the Internet shortly. This big resolution will have lots of pros and cons in the technology. The positive is that we can track all our belongings through web applications. It will help in the productivity of agriculture fields and, at the same time, will be helpful for the security of endangered species. IoT also helps in enhancing a country's security. The way IoT has emerged in recent times shows a new industrial revolution. But connecting every device, from small to big, to a network has led to a broader scope for attacks. This paper focuses on one of the IoT-specific, which is Hatchetman attack. The attack potentially decreases the PDR and increases the error messages in the network. This makes the network unstable and it tries to reconfigure itself by resetting the trickle timer. 
\par In the future, we plan to implement different IoT-specific attacks and generate a  multi-label attack dataset to detect the network's behavior. We also plan to showcase the attack's impact, where the attackers can coordinate with each other to make the attack more impactful. Coordinated attacks drastically impact the network, and the attacker nodes are not easily identifiable. 
% \bibliographystyle{IEEEtran}
% \bibliography{mybib}

% \end{IEEEbiography}



\end{document}